\documentclass[twocolumn]{aastex631}
\usepackage{float}
\usepackage{amsmath}

\begin{document}

\title{New Potential Ultra-compact X-ray Binaries for Space-based Gravitational Wave Detectors From Low-Mass Main-Sequence Companion Channel}

\author{Minghua Chen}
\affiliation{Xinjiang Astronomical Observatory, Chinese Academy of Sciences, Urumqi 830011, China; liujinzh@xao.ac.cn}
\affiliation{University of Chinese Academy of Sciences, Beijing 100049, People's Republic of China }


\author{Jinzhong Liu}
\affiliation{Xinjiang Astronomical Observatory, Chinese Academy of Sciences, Urumqi 830011, China; liujinzh@xao.ac.cn}
\affiliation{University of Chinese Academy of Sciences, Beijing 100049, People's Republic of China }



\begin{abstract}

We investigate the formation and evolution of Ultra-Compact X-ray Binaries (UCXBs) using the COMPAS binary evolution code, starting from the Zero Age Main Sequence (ZAMS). Focusing on the low-mass MS companion channel, we simulate gravitational wave (GW) signals from UCXBs with LEGWORK and evaluate their detectability by space-based observatories such as Taiji and TianQin. By incorporating signal-to-noise ratio (SNR) calculations with a threshold of $\text{SNR} > 5$, we provide a realistic framework to assess the detectability of the GW source.
Our analysis suggests that the Milky Way currently hosts \(7{-}32\) observable UCXBs from the MS companion channel. Taiji or LISA alone could detect \(1{-}6\) sources over an 8-year observation period, \textbf{while TianQin, due to its high-frequency sensitivity, contributes to detecting systems with extremely short orbital periods and can also detect \(1{-}4\) sources. }
Comparison with sensitivity curves validates UCXBs as detectable GW sources, particularly at greater Galactic distances.
This study improves our understanding of the evolution of UCXBs and their role as GW sources. By integrating population synthesis, SNR-based analyses, and observational data, we establish UCXBs as significant targets for GW astronomy, paving the way for future missions and theoretical studies of compact binary systems.

\end{abstract}



\section{Introduction} \label{sec:intro}
Ultracompact X-ray binaries (UCXBs) are low-mass X-ray binaries (LMXBs) with ultra-short orbital periods typically less than 60 minutes, consisting of a neutron star (NS) or black hole (BH) accretor and a hydrogen-poor donor star 
 \citep{1982ApJ...254..616R,1986ApJ...304..231N}. UCXBs play a vital role in exploring the evolution of binary stars \citep{2021MNRAS.503.3540C,2023pbse.book.....T}, the structure of accretion disks \citep{2013ApJ...768..184H} and the behavior of accreting millisecond pulsars especially the development of spider pulsars \citep{2023ApJ...956L..39Y}. To date, approximately 20 sources in our Galaxy have been confidently confirmed as UCXBs based on precise orbital period measurements \citep{2023A&A...677A.186A}. In most confirmed UCXBs, NS act as the accretor. The source X9 in the globular cluster 47 Tucanae was thought to be a \textbf{possible} BH-UCXB candidate, whose nature remains undetermined \citep{2017MNRAS.467.2199B,2017HEAD...1610817C,2023A&A...677A.186A}.
 
 UCXBs in the Galactic field are believed to result from binary star evolution through \textbf{four channels}: \textbf{(i) a white dwarf (WD) transferring matter through the Roche-lobe overflow (RLOF) to a NS/BH} \citep{1975MNRAS.172..493P,2002A&A...388..546Y,2004ApJ...616L.139W,2017MNRAS.470L...6S}. (ii) a helium (He) star donor providing He-rich \textbf{material }to fill its Roche lobe \citep{1986A&A...155...51S,2008AstL...34..620Y,2021MNRAS.506.4654W}. \textbf{(iii) a low-mass main-sequence (MS) star near the end of hydrogen core burning filling its Roche lobe with initial orbital periods shorter than the bifurcation period}\citep{1986ApJ...304..231N,2002ApJ...565.1107P,2005A&A...431..647V,2021MNRAS.503.2776Y}.  \textbf{(iv) the accretion-induced collapse (AIC) of ONe WDs in binaries can also contribute to the formation of UCXBs} \citep{2013A&A...558A..39T,2020RAA....20..135W,2023MNRAS.521.6053L}. These evolutionary scenarios highlight the profound role of UCXBs in studying accretion physics, angular momentum loss, and the interplay between gravitational and nuclear processes \citep{2013ApJ...768..184H,2021MNRAS.503.3540C,2023pbse.book.....T}. Beyond their relevance to compact binary systems, UCXBs also offer broader insights into astrophysics, providing a window into stellar evolution under extreme conditions \citep{2013ApJ...768..184H}.
 
The advent of gravitational wave (GW) astronomy has revolutionized our understanding of the universe, with space-based observatories poised to open a new frontier in low-frequency GW detection \citep{2017arXiv170200786A}. UCXBs, with their short orbital periods, are key targets for space-based GW detectors such as LISA \citep{2023LRR....26....2A}, Tianqin \citep{2016CQGra..33c5010L}, and Taiji \citep{2020IJMPA..3550075R}. During their inspiraling phases, UCXBs emit low-frequency GW signals, providing crucial insights into binary interactions, envelope ejection, and compact binary evolution \citep{2013A&ARv..21...59I,2000A&A...360.1011N}. Studying UCXBs through these observatories not only enhances our understanding of these systems but also enriches the broader field of astrophysics by revealing the fundamental physics governing their behavior \citep{2013ApJ...768..184H,2023LRR....26....2A}.

However, the GW radiation from UCXBs, combined with the low-mass MS companion channel, are not well studied. We focus on this channel in the current study, where UCXBs originate from low-mass MS stars near the end of hydrogen core burning. This channel, despite its potential significance, has been underexplored compared to other formation pathways. we simulate the evolutionary pathways of potential UCXBs from zero-age main-sequence (ZAMS) stars and model their corresponding GW signals. The simulated companion stars in our study have notably low masses, aligning well with previous works in the field \citep{2020ApJ...900L...8C}. By comparing the simulated signals with the detection capabilities of space-based GW observatories, and incorporating observational data from confirmed UCXB sources, this study validates the feasibility of detecting UCXBs through GW astronomy and underscores their importance as astrophysical laboratories for extreme conditions.

\section{Gravitational wave radiation from UCXBs} \label{sec:style}

UCXBs are characterized by their very short orbital periods, typically less than one hour, and involve complex mass transfer processes during their evolution. Due to their tight orbits and high mass densities, UCXBs are significant sources of GWs \citep{2010NewAR..54...87N,2020ApJ...900L...8C}.

\subsection{GW Emission Characteristics in UCXBs}
The characteristics of GW emission in these systems can be understood through the frequency, strain amplitude, and radiation power. 

 For an circular binary system, the GW frequency depends on the orbital period. It can be expressed as:

\begin{equation}
  f_{\text{gw}} = \frac{2}{P_{\text{orb}}} .
\end{equation}

By analyzing the frequency, we can further deduce the corresponding strain. The strain amplitude can be expressed as:

\begin{equation}
  h_0 = \frac{2 (G M_c)^{5/3}}{c^4} 
  \left( \frac{\pi f_{\text{gw}}}{D} \right)^{2/3} .
\end{equation}
where \(M_c = \frac{(m_1 m_2)^{3/5}}{(m_1 + m_2)^{1/5}}\) is the chirp mass and \( D \) is the distance to the source.

The radiation power $P_{gw}$ emitted by UCXB due to GWs, which significantly impact the orbital period, mass transfer rate, and energy \citep{2010NewAR..54...87N}, is given by:

\begin{equation}
P_{\text{gw}} = \frac{32}{5} \frac{(2\pi)^{8/3}}{G^{5/3} c^5} 
\frac{(m_1 m_2)^2}{(m_1 + m_2)^{1/3} P_{\text{orb}}^{10/3}} .
\end{equation}

These relationships highlight how the orbital dynamics of UCXBs drive GW emission, with shorter orbital periods producing higher-frequency and higher-amplitude signals.

\subsection{Mass Transfer}

In UCXBs, mass transfer occurs through Roche lobe overflow. As the donor star fills its Roche lobe, material is transferred to the compact object through an accretion disk \citep{2012A&A...537A.104V}. The mass transfer rate ($\dot{M}_d$) is determined by the formula:

\begin{equation}
  \dot{M} = \frac{2 \dot{J}}{J \left( \frac{\Delta R_1}{\Delta m_1} - \frac{\Delta R_2}{\Delta m_2} \right)} ,
\end{equation}
where \(J\) is the orbital angular momentum, and \(\frac{\Delta R_1}{\Delta M_1}\), \(\frac{\Delta R_2}{\Delta M_2}\) describe the sensitivity of the donor and Roche lobe radii to mass changes, respectively. 

Initially, the shrinking orbit driven by GW radiation causes a rapid increase in the mass transfer rate. \textbf{The orbital evolution of UCXB systems under the influence of GW radiation during the mass transfer process exhibits significant complexity. In general, GW radiation acts to reduce the orbital separation by extracting angular momentum from the system. However, in certain cases, if the mass transfer rate is sufficiently high, the system's total angular momentum may increase, leading to an increase in orbital separation \citep{2001A&A...375..890N,2014LRR....17....3P,2022MNRAS.515.2725G}. This expansion, in turn, may lead to a reduction in the mass transfer rate and a potential decrease in the efficiency of GW radiation emission.}

\subsection{Energy Loss and Orbital Decay}
The energy carried away by GWs leads to orbital decay. This continuous energy loss through GW radiation is a critical factor in the orbital decay process \citep{2012A&A...537A.104V}. The rate of energy loss due to gravitational radiation is given by:
\begin{equation}
\frac{dE}{dt} = - \frac{32}{5} \frac{G^4}{c^5} 
\frac{(m_1 m_2)^2 (m_1 + m_2)}{a^5} .
\end{equation}
where \(a\) is the orbital separation. As the separation decreases, the rate of energy loss increases, leading to faster orbital decay. The corresponding rate of orbital period change is given by:

\begin{equation}
\frac{dP_{\text{orb}}}{dt} = 
-\frac{96 \pi G^{5/3}}{5 c^5} 
\left( \frac{2 \pi}{P_{\text{orb}}} \right)^{5/3} 
\frac{m_1 m_2}{(m_1 + m_2)^{1/3}}.
\end{equation}

These equations demonstrate the close interplay between energy loss and orbital evolution, which ultimately leads to mergers that produce compact remnants such as neutron stars or black holes \citep{2021MNRAS.506.4654W}.

\subsection{Detector Capability}
UCXBs emit GWs within the sensitivity range of space-based GW observatories such as Taiji, TianQin, and LISA. These observatories are designed to detect low-frequency GWs.

The detectability of GW signals can be quantified by the signal-to-noise ratio (SNR):

\begin{equation}
\text{SNR} = \sqrt{4 \int_0^\infty \frac{|\tilde{h}(f)|^2}{S_n(f)} \, df},
\end{equation}
where \(\tilde{h}(f)\) is the Fourier transform of the GW signal, and \(S_n(f)\) is the noise spectral density of the detector. Observations over longer durations enhance the SNR as:

\begin{equation}
\text{SNR} \propto \sqrt{T}.
\end{equation}

Taiji and LISA, both optimized for low-frequency GWs, excel in detecting the early inspiral phases of UCXBs, while TianQin’s geocentric orbit enhances its sensitivity to systems nearing merger. Together, these observatories provide complementary insights into the full evolutionary pathway of UCXBs, covering both early inspiral and late merger stages.

Typically, an SNR greater than 5 or 8 is considered to indicate a credible detection.

The above formulas and parameters will serve as the foundation for our model to simulate the evolution of UCXBs and their GWs, allowing us to better evaluate the capability of space-based GW detectors to detect GWs from UCXBs.

\section{Population Synthesis and Gravitational Wave Simulation Model} \label{sec:floats}

\subsection{Population Synthesis Model}
To reliably simulate the GW radiation of UCXBs, we employ the Compact Object Mergers: Population Astrophysics and Statistics (COMPAS) code \citep{2022ApJS..258...34R}. This tool is specifically designed to model the evolution of compact object binaries, starting from the ZAMS and progressing through critical stages, including supernova kick, mass transfer, CE evolution, RLOF, and the formation of compact objects.

A key feature of COMPAS is its ability to rapidly estimate the evolution of binary systems using approximate models, which allows for the exploration of large parameter spaces. This approach, while computationally efficient, may limit the detailed physical accuracy of certain stages, such as the CE and RLOF phases, which involve complex and highly non-linear processes.

The CE phase plays a crucial role in the evolution of UCXBs, where the primary star expands and engulfs its companion, forming a shared gas envelope. This phase induces a significant loss of angular momentum, driving the binary components closer together. In COMPAS, angular momentum loss mechanisms such as magnetic braking and GW radiation are incorporated using parameterized models \citep{1964PhRv..136.1224P}. The energy balance in the CE phase is described by the $\alpha_{\rm CE}$-formalism, which simplifies the process but captures its overall effects on binary evolution.

RLOF is another key phase in UCXBs' evolution, during which mass transfer occurs as the donor star fills its Roche lobe. The Roche lobe radius (\(R_L\)) is approximated by the following relation \citep{1990A&A...236..385K}:

\begin{equation}
\frac{R_L}{a} \approx \frac{0.49 q^{2/3}}{0.6 q^{2/3} + \ln(1 + q^{1/3})},
\end{equation}

where \(a\) is the orbital separation and \(q\) is the mass ratio of the binary system, defined as \(q = \frac{m_2}{m_1}\).

Building on its success in modeling the binaries BH-BH, BH-NS, and NS-NS \citep{2022ApJ...937..118W}, COMPAS is extended here to simulate the formation and evolution of UCXBs. Using large-scale BPS, we estimate the population properties of UCXBs to evaluate their detectability by space-based GW detectors. This approach highlights the statistical trends of UCXBs rather than their detailed evolution, providing a comprehensive framework for understanding their role as GW sources.

\textbf{In this context, we employed the COMPAS binary evolution code to simulate a comprehensive parameter space, to investigate the formation and evolution of UCXBs. The grid spans the following ranges: the initial accretor mass (\(M_1\)) ranges from 6–12 \(M_\odot\) with a step size of 0.1 \(M_\odot\), and the initial donor star mass (\(M_2\)) spans 0.4–1.2 \(M_\odot\) with a step size of 0.01 \(M_\odot\). The initial orbital radius (\(R\)) ranges from 0.5–10 AU in increments of 0.5 AU, and the system metallicity is varied between 0.001 and 0.01 with a step size of 0.001. This parameter grid ensures a thorough exploration of UCXBs’ potential formation pathways and evolutionary outcomes. All simulations assume circular orbits, consistent with the low eccentricity typically observed in UCXBs due to tidal forces.}

\textbf{Starting from a ZAMS binary system in the COMPAS code, the evolution to an UCXB involves several key stages. Initially, the system consists of two close binary stars on the main sequence, with a significant mass difference between them (e.g., $M_1$/$M_2$ $>$ 8). The more massive primary star evolves into a red giant, losing mass through stellar winds. Despite this mass loss, the core of the primary may undergo a supernova explosion, leading to the formation of a compact object, such as a neutron star or black hole. Meanwhile, the donor star remains on the main sequence and undergoes mass transfer to the compact object, typically through RLOF, resulting in significant angular momentum loss. This process may trigger CE evolution, and the CE phase is often unstable, leading to additional mass loss and further orbital contraction . The orbital period decreases, eventually forming a LMXB. As the orbit becomes extremely tight, the efficiency of mass transfer from the donor to the compact object increases, leading to the formation of a UCXB from the MS companion channel. According to the model proposed by  \citet{2022ApJS..258...34R}, a star is assumed to experience an electron-capture supernova (ECSN) if its He core mass falls within the range of 1.6 to 2.25 solar masses, while stars with He core masses exceeding 2.25 solar masses are expected to undergo core-collapse supernovae (CCSNe). These formation criteria are crucial for understanding the subsequent evolution of UCXBs.}

\subsection{Gravitational Wave Simulation Model}
To simulate the GW signals of UCXBs, we employ the LISA Evolution and Gravitational Wave Orbit Kit (LEGWORK) \citep{2022ApJS..260...52W}. This Python-based toolkit enables simulations of orbital dynamics, GW signal prediction, and detector sensitivity. LEGWORK natively supports the LISA and Tianqin detectors \citep{2019CQGra..36j5011R}, and we adapt it to the Taiji model by replacing the sensitivity parameters of LISA with those specific to Taiji \citep{2020IJMPA..3550075R}.

The sensitivity of the Taiji detector can be described by the following equation, \textbf{similar to that of LISA}, which accounts for noise contributions from the optical metrology system, single-test mass acceleration, and galactic confusion noise:

\begin{equation}
\begin{aligned}
S_n(f) &= \frac{1}{L^2} \left( P_{\text{OMS}}(f) + 2 \left( 1 + \cos^2\left( \frac{f}{f_*} \right) \right) \frac{P_{\text{acc}}(f)}{(2 \pi f)^4} \right) \\
\quad \times \frac{1}{R} + S_c(f),
\end{aligned}
\end{equation}

where \(L \) is the detector arm length, \(f_* = \frac{c}{2\pi L} \) is the transfer frequency, \(P_{\text{OMS}}(f)\) is the optical metrology noise, \(P_{\text{acc}}(f)\) is the single-test mass acceleration noise, and \(S_c(f)\) is the galactic confusion noise. 

For the response function \( R \), we adopted an approximate estimation to simplify the calculations while maintaining sufficient accuracy. The response function describes the detector's sensitivity to GW signals at different frequencies, and its approximate form is given by:

\begin{equation}
R= \frac{3}{10} \left( 1 + 0.6 \left( \frac{f}{f_*} \right)^2 \right),
\end{equation}

\begin{figure}
\centering
\includegraphics[scale=0.3]{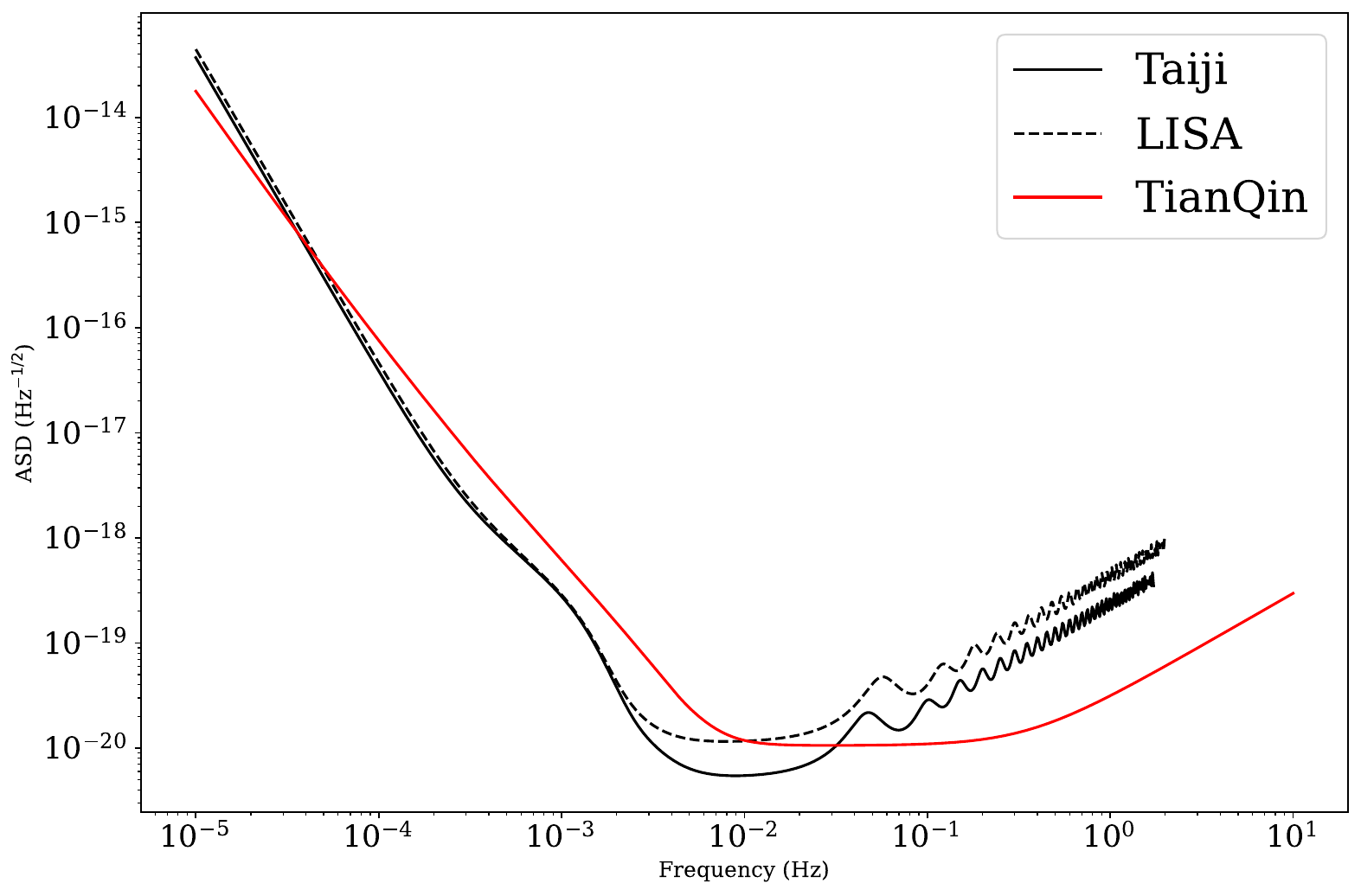}
\caption{The figure presents the sensitivity curves of Taiji, LISA, and TianQin detectors. The solid black line represents the sensitivity of the Taiji detector, the dashed line represents the sensitivity of the LISA detector, and the red line represents the sensitivity of the TianQin detector. All curves are plotted as Amplitude Spectral Density (ASD) versus frequency.}
\label{fig:1}
\end{figure}

Using the parameters, we implement the Taiji sensitivity curve into LEGWORK and compare it with the LISA sensitivity curve. As shown in Fig. \ref{fig:1}, Taiji demonstrates improved sensitivity in specific frequency ranges compared to LISA, particularly at lower frequencies, making it better suited for detecting inspiral stages of UCXBs.

The Tianqin sensitivity curve is also implemented in LEGWORK, enabling a comparison of its performance against Taiji and LISA. Tianqin is optimized for higher-frequency signals, with its sensitivity curve given by:

\begin{equation}
\begin{aligned}
S_n(f) &= \frac{10}{3 L^2} \left( \frac{4 S_a}{(2 \pi f)^4} \left( 1 + \frac{1 \times 10^{-4}}{f} \right) + S_x \right) \\
\quad \times \left( 1 + 0.6 \left( \frac{f}{f_*} \right)^2 \right) + S_c(f),
\end{aligned}
\end{equation}

LEGWORK has been successfully applied to simulate GWs from compact object binaries, including binary black holes (BBHs) and neutron star binaries (NSBs) \citep{2022ApJ...937..118W}. In this study, we extend its application to UCXBs, focusing on modeling their GW characteristics and evaluating their detectability by comparing these signals with detector sensitivity curves.

Additionally, LEGWORK's capability to incorporate observational data provides a unique opportunity to align theoretical models with real astrophysical systems. By integrating data from confirmed UCXBs, we enhance the accuracy of simulated GW signals, ensuring they reflect realistic astrophysical parameters. This approach not only validates the robustness of LEGWORK but also provides a framework for testing the detectability of UCXBs under varying observational scenarios. Such simulations are essential for advancing the design and optimization of future GW detection missions and for deepening our understanding of UCXBs as key sources in the low-frequency GW regime.

\section{Observational data on UCXBS} \label{sec:cite}

The study of UCXBs began in the late 1970s with the HEAO-1 satellite \citep{1984ApJS...56..507W}, which pioneered the identification of X-ray properties in compact binary systems. Since then, continuous advancements in observational techniques have facilitated the discovery of UCXBs using instruments such as RXTE \citep{1996SPIE.2808...59J}, Chandra \citep{2000SPIE.4012....2W}, XMM-Newton \citep{2001A&A...365L...1J}, and Swift \citep{2004ApJ...611.1005G}.

RXTE, with its advanced timing capabilities and broad energy range, played a crucial role in studying UCXBs. Notably, it facilitated the discovery of accreting millisecond X-ray pulsars such as XTE J1751-305, enabling precise mass measurements \citep{2002ApJ...575L..21M}. These findings significantly advanced our understanding of compact object physics and binary evolution.

The Chandra X-ray Observatory, launched by NASA in 1999, is among the most advanced X-ray telescopes, offering unparalleled imaging and spectroscopic capabilities. Chandra has significantly contributed to the study of UCXBs by resolving fine details of X-ray emissions and structures. For instance, observations of 4U 1916-05 revealed X-ray dips that provided critical insights into the system's geometry and dynamics \citep{2006ApJ...646..493J}. Additionally, Chandra’s data on 47 Tuc X9 identified it as a potential black hole UCXB, deepening our understanding of these rare systems \citep{2005ApJ...625..796H}.

XMM-Newton, the X-ray Multi-Mirror Mission launched by the European Space Agency (ESA) in 1999, has significantly advanced the study of UCXBs. Its high sensitivity and spectroscopic capabilities have yielded critical data on systems such as IGR J17062-6143 \citep{2019MNRAS.488.4596H} and 2S 0918-549 \citep{2014MNRAS.442.2817K}, offering new insights into their emission mechanisms and orbital properties.

Swift has been instrumental in discovering and confirming new UCXB systems through its combined X-ray and optical observational capabilities. By detecting X-ray bursts and corresponding optical changes, Swift has established the nature of several candidates. Notably, it observed an outburst from Swift J0911.9-6452, enabling a detailed study of this source. Furthermore, Swift identified Swift J1756.9-2508 as a confirmed UCXB \citep{2007ApJ...668L.147k}, providing crucial insights into the nature and evolution of these compact systems.

Precise distance measurements are essential for studying GWs from UCXBs. The Gaia mission, launched by the ESA in 2013, has revolutionized astrometry by providing accurate positions, distances, and proper motions for over a billion stars. Gaia’s data significantly improves the distance estimates for UCXBs such as 2S 0918-549, 4U 1543-624 \citep{2007ApJ...668L.147k}, and 47 Tuc X-9 \citep{2018ApJ...867..132C}. In our study, we references data from Gaia DR2, utilizing the distances of some known UCXBs to calculate and simulate GW emissions.

Although we acknowledge that there is significant uncertainty regarding the evolutionary channels of UCXBs, for example, XTE J1751-305 could potentially have evolved from an ONe WD + He WD system \citep{2023MNRAS.521.6053L}. We chose five sources with a broad range of orbital periods and distances. \textbf{Among the 20 known UCXBs, previous studies have investigated the formation and evolution of 15 of them in detail using multi-wavelength observational data \citep{2014MNRAS.442.2817K,2015ApJ...798..117P,2019MNRAS.488.4596H,Yuan_2025}, providing some insight and potential constraints for theoretical studies on the formation and evolution of UCXBs \citep{2020ApJ...900L...8C,2023ApJ...944...83Q,2024ApJ...961..110Q}. However, for the remaining 5 systems, there is still considerable uncertainty and ongoing discussion regarding their possible evolutionary channels. Considering the relative contributions of different UCXBs formation channels, and based on the relatively stable X-ray emission characteristics observed in these 5 UCXBs \citep{2021ApJ...911..123L,2021ApJ...920..142H}, this study tentatively proposes that they might be candidates for the low-mass MS companion channel in UCXB formation.}

\begin{table*}
\footnotesize
\caption{Information of selected UCXBs}
\label{tab:1}
\tabcolsep 6pt 
\begin{tabular*}{\textwidth}{@{\extracolsep{\fill}} p{0.3cm} c c c c c c c @{}}
\hline
ID & Name & RA & DEC & D (Kpc) & $P_{\text{orb}} (min)$ & $M_1$ ($M_\odot$) & $M_2$ ($M_\odot$) \\
\hline
1  & 4U 1543-624  & 15:47:54.69 & -62:34:05.4 & $3.3^{+6.5}_{-1.4}$ & $18.2^{+0.1}_{-0.1}$  & $1.4$ & $0.0432$ \\
2  & 4U 1626-67   & 16:32:16.79 & -67:27:39.3 & $3.5^{+2.3}_{-1.3}$ & $41.538^{+0.002}_{-0.002}$ & $1.4$ & $0.0189$ \\
3  & XTE J1751-305 & 17:51:13.49 & -30:37:23.4 & 8.5 & $42.42235^{+0.00006}_{-0.00006}$ & $1.4$ & $0.0185$ \\
4  & XTE J0929-314 & 09:29:20.19 & -31:23:03.2 & 10 & $43.57910^{+0.00005}_{-0.00005}$ & $1.4$ & $0.0181$ \\
5  & 4U 1916-053  & 19:18:47.87 & -05:14:17.09 & $7.6^{+0.4}_{-0.4}$ & $49.75^{+0.16}_{-0.16}$ & $1.4$ & $0.0158$ \\
\hline

\end{tabular*}
\begin{minipage}{\textwidth}
    \footnotesize
    \textbf{Note.} The $M_1$ is assumed to be the mass of a standard neutron star (fixed at $1.4 M_\odot$), while the $M_2$ is estimated using the formula by Chen et al. \citep{2020ApJ...900L...8C}.
    
    \textbf{References.} Distances and orbital periods of the selected UCXBs are derived from direct observational data, as reported in previous studies \citep{1981ApJ...244.1001M,2002ApJ...575L..21M,2002ApJ...576L.137G,1982ApJ...253L..67W,2006ApJ...646..493J}.
\end{minipage}
\end{table*}

To present the observational data, Table \ref{tab:1} provides key observational parameters for five selected UCXBs. The listed parameters are essential for our analysis and simulations, offering insights into the physical and orbital properties of these systems. Given the influence of tidal forces, UCXBs typically exhibit near-circular orbits with very low eccentricity. Thus, the calculations presented in the table assume circular orbits (\( e = 0 \)) \citep{2000A&A...360.1011N}.

\section{Results and discussion} \label{sec:cite}

\begin{figure}[h!]
\centering
\includegraphics[scale=0.2]{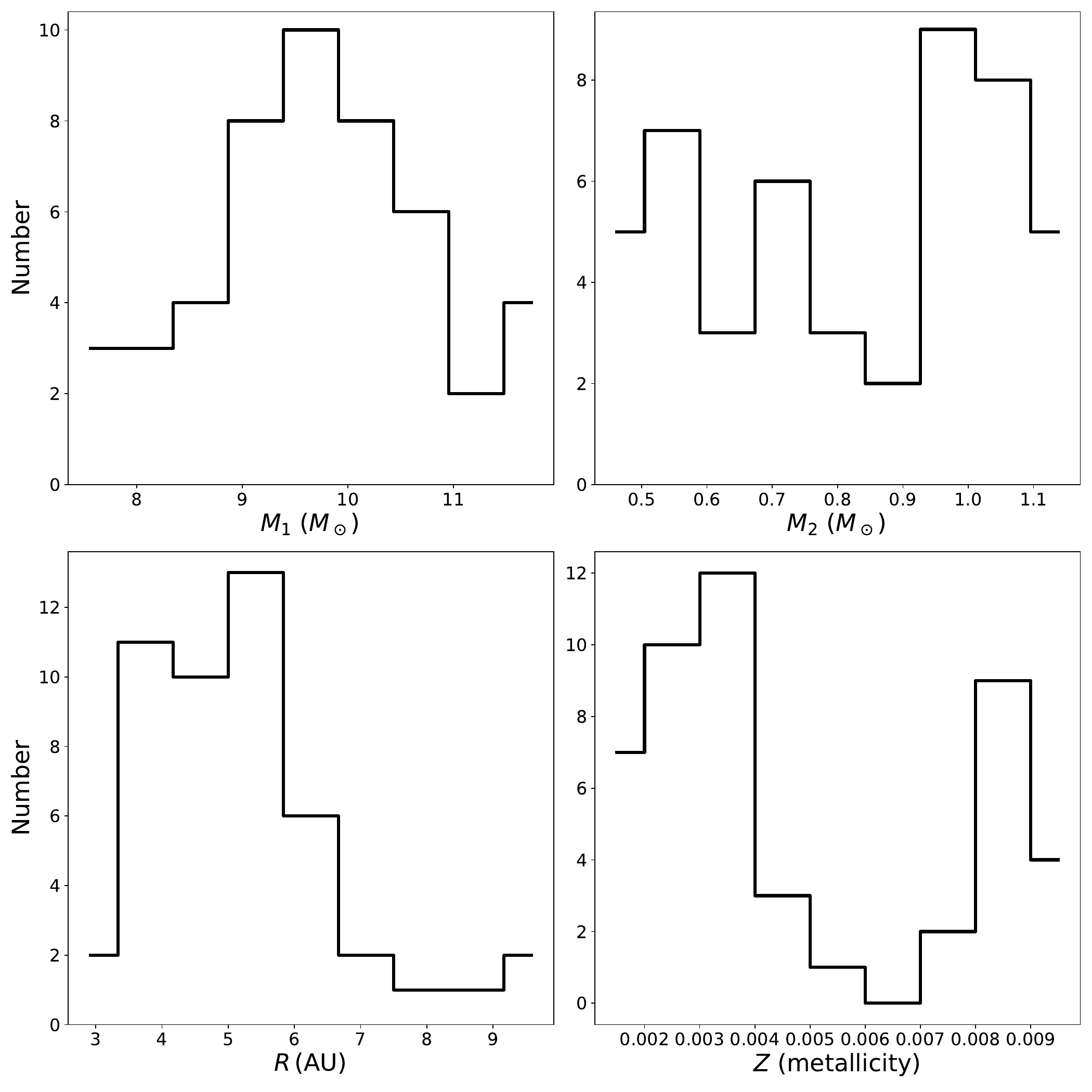}
\caption{Distributions of the initial conditions for the simulated UCXB systems. The panels show the distributions of the $M_1$, $M_2$, $R$, and metallicity.}
\label{fig:2}
\end{figure}

From the ZAMS, we evolved the binary systems and identified 48 potential UCXB systems.\textbf{ We also attempted to enlarge the samples of primordial ZAMS binaries in COMPAS binary evolution code, but the results indicated that the simulation had already reached convergence. Therefore, variations in the initial sample size on the order of millions would not significantly impact the evolved parameters of UCXBs in our simulations.} Analysis of their initial conditions reveals distinct trends across parameters, including accretor and donor star masses, orbital radii, and metallicity distributions, as illustrated in Figure \ref{fig:2}.

The accretor mass (\(M_1\)) distribution is centered around \(9.5 M_\odot\), with most systems falling between \(8 M_\odot\) and \(11 M_\odot\), suggesting that moderately massive accretors are more conducive to UCXB formation. Donor star masses (\(M_2\)) exhibit a broad distribution, with prominent clustering around \(0.7 M_\odot\) and \(1.0 M_\odot\), reflecting a wide range of viable mass transfer rates that contribute to the diversity of UCXBs. The initial orbital separations (\(R\)) are predominantly concentrated between \(4–6 \, \text{AU}\), highlighting the importance of moderate separations for UCXB formation. Excessively wide orbits fail to initiate efficient mass transfer, while overly tight orbits risk premature mergers. This distribution underscores the delicate balance required between orbital separation and stellar evolutionary timescales. The metallicity distribution skews toward lower values, with most systems around \(0.002\). Low-metallicity environments favor UCXB formation by reducing stellar wind mass loss, enabling binaries to retain more mass. This stabilizes mass transfer and enhances the likelihood of binaries shrinking into the ultra-compact phase.

\begin{figure}[h!]
\centering
\includegraphics[scale=0.2]{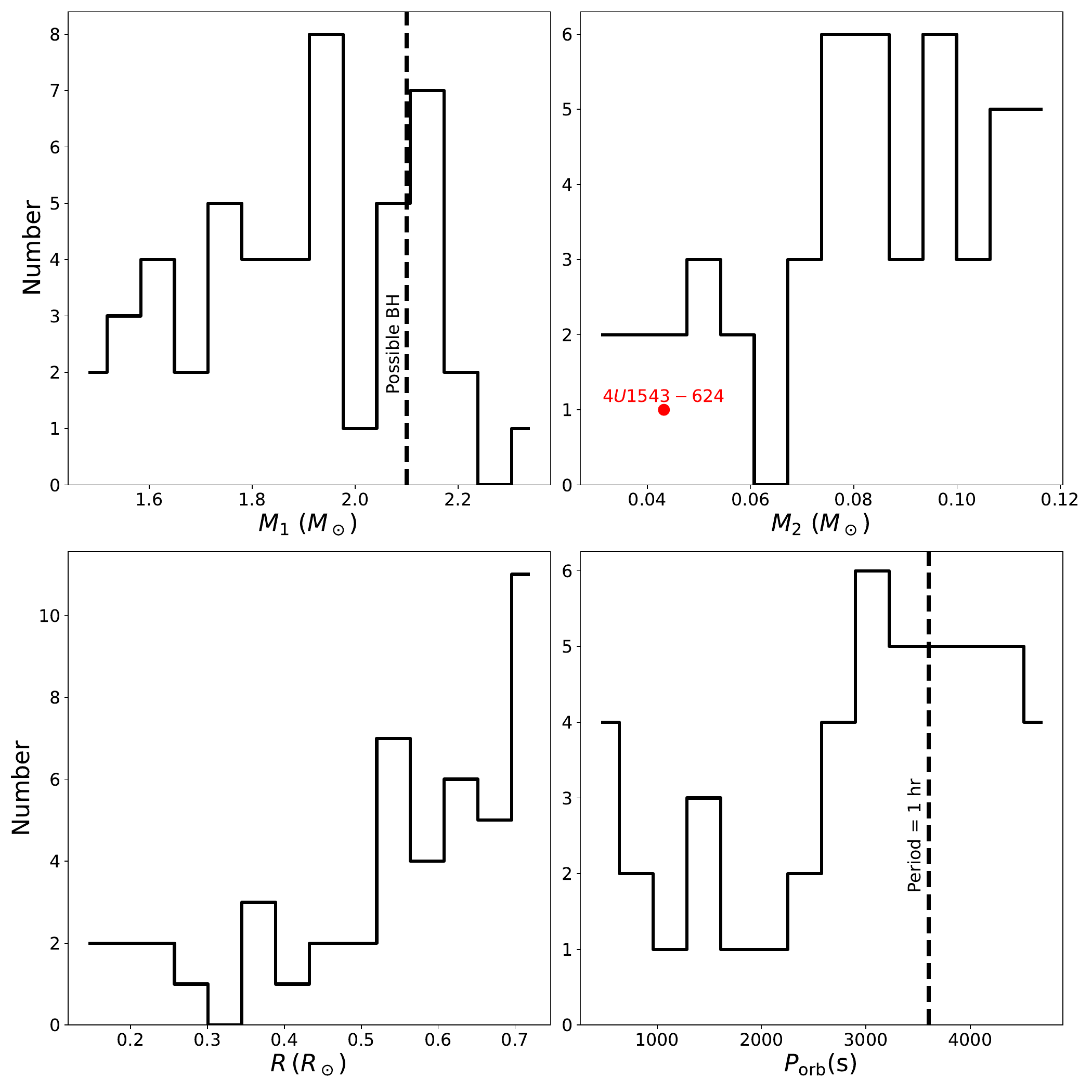}
\caption{Distributions of the evolved parameters for the simulated UCXB systems, showing $M_1$, $M_2$, $R$, and orbital period. The dashed lines indicate the assumed possible black hole and critical orbital period of 1 hour.}
\label{fig:3}
\end{figure}

Following the evolution of the systems, we analyzed the final parameters of the UCXBs. These systems exhibited significant changes in their accretor and donor masses, orbital radius, and periods. By examining the evolved conditions of the UCXBs, we observed several key trends, as shown in Figure \ref{fig:3}.

The figure illustrates the distributions of evolved parameters for simulated UCXB systems, highlighting the final stages of their evolution in terms of $M_1$, $M_2$, $R$, and orbital period. These distributions provide insight into the characteristics of UCXBs as they evolve into their ultra-compact phase.

For the $M_1$ distribution, most evolved accretor masses lie between $1.6 M_{\odot}$ and $2.3 M_{\odot}$. The dashed line at $2.1 M_{\odot}$ marks the threshold used to distinguish neutron stars from possible black holes. Systems with accretor masses exceeding $2.1 M_{\odot}$ are likely candidates for black hole UCXBs, indicating that some systems may evolve into black holes rather than neutron stars. This distinction underscores the diversity of outcomes in the evolutionary pathways of UCXBs.

The $M_2$ distribution shows that the majority of donor stars have masses below $0.1 M_{\odot}$, consistent with significant mass loss over time. Such low masses reflect advanced stages of donor evolution, where prolonged mass transfer has nearly depleted the donor star, leaving it with minimal mass. This feature is characteristic of UCXBs nearing the ultra-compact phase. Previous research has shown that the ultra-compact phase resulting from the NS + MS companion evolutionary channel begins with a companion star mass around $0.1 M_{\odot}$ \citep{2020ApJ...900L...8C}.

The third panel shows the distribution of orbital radius ($R$), where most systems have compact orbits, ranging from $0.5 R_{\odot}$ to $0.7 R_{\odot}$. These tight radius are driven by continuous mass transfer and GW radiation, which gradually shrink the orbits. Such compact orbits are a defining trait of UCXBs, facilitating the conditions for sustained mass transfer and orbital tightening.

In the orbital period distribution, most systems exhibit periods under 1 hour, which is typical for UCXBs. The dashed line at 1 hour indicates the threshold for confirmed UCXBs, though we extend the classification to systems with periods up to 90 minutes to encompass more potential UCXB candidates. Some extreme cases show periods as short as 10 minutes, illustrating advanced stages of orbital contraction due to GW radiation. These distributions capture the wide range of evolutionary paths observed in UCXBs, validating the model’s ability to reflect the diversity of UCXB systems found in nature.

\subsection{The Evolution of UCXBs}

\begin{figure}[H]
\centering
\includegraphics[scale=0.2]{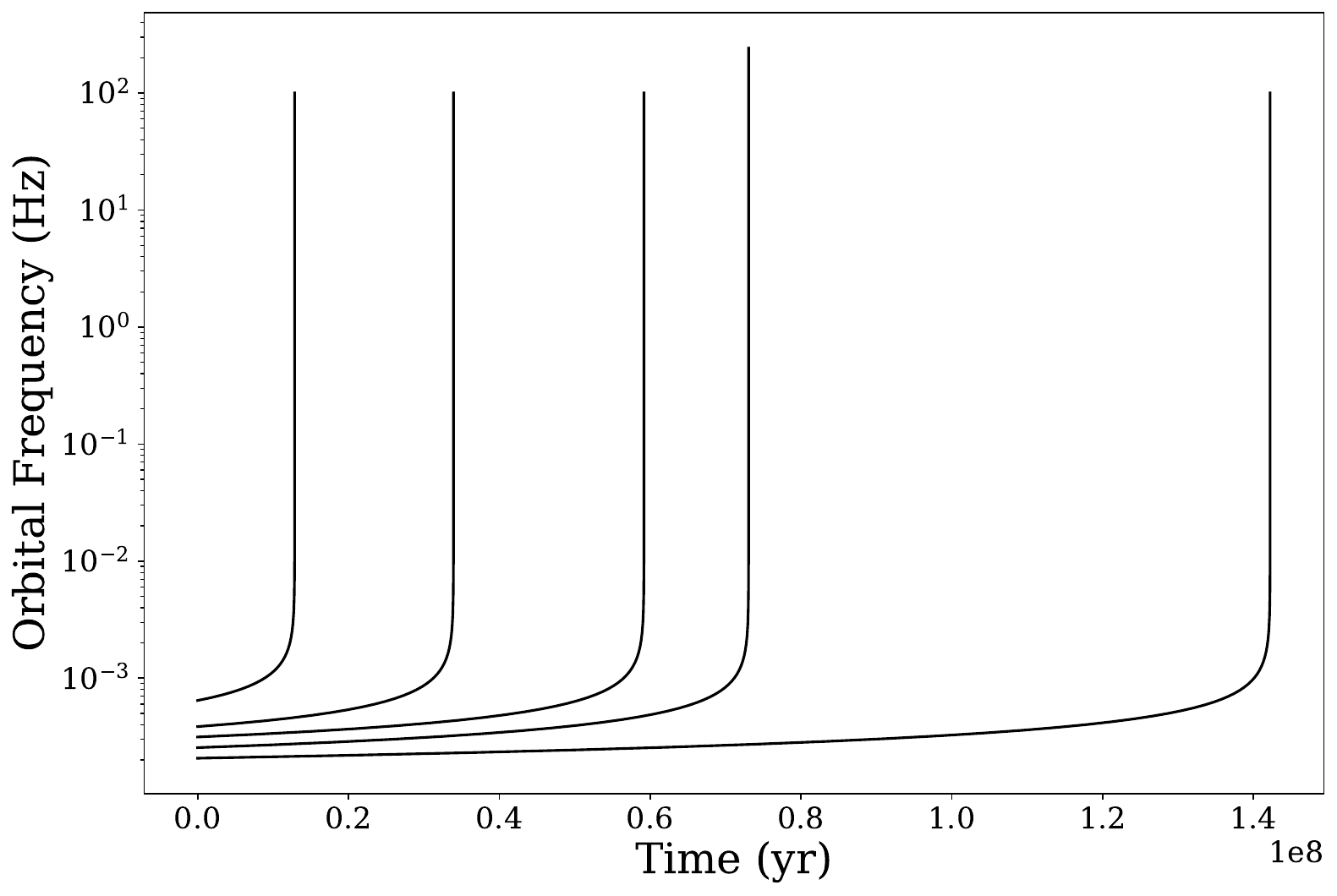}
\caption{Orbital frequency evolution of the UCXB system, generated using LEGWORK}
\label{fig:4}
\end{figure}

In this section, we explore the detailed orbital evolution of UCXB systems, focusing on the results generated using the LEGWORK tool. The simulation results, as illustrated in Figure \ref{fig:4}, demonstrate a gradual increase in the orbital frequency over time, driven primarily by GW radiation,  This steady increase reflects the stable shrinking of the binary’s orbit as it evolves toward the ultra-compact phase.

The later stages of evolution show a sharp rise in orbital frequency, underscoring the accelerating effect of GW emission on orbital tightening. This rapid contraction phase marks a critical period where the binary emits intense GW signals, making it particularly significant for observational studies.

From our simulations, the lifetimes of UCXBs formed through the MS channel are approximately 
$50{-}150 \, \mathrm{Myr}$, defined as the period during which the systems emit detectable GW signals. This result is consistent with Wang et al. \citep{2021MNRAS.506.4654W}, who reported similar lifetimes for UCXBs from MS channel. These systems evolve more gradually due to their lower initial mass transfer rates, allowing for a longer detection window and sustained GW emission. Despite rapid changes near the end of their lifetimes, the systems remain dynamically stable for the majority of their evolution, ensuring that UCXBs are long-lived sources of GWs. This stability and prolonged GW emission make UCXBs particularly valuable targets for space-based GW detectors, which can track their signals across different evolutionary stages, providing insights into their dynamics and the underlying physics of compact binaries.

\subsection{Simulated Gravitational Wave Characteristics of UCXBs}

\subsubsection{Properties of the Gravitational Waves}

\begin{figure}[H]
\centering
\includegraphics[scale=0.3]{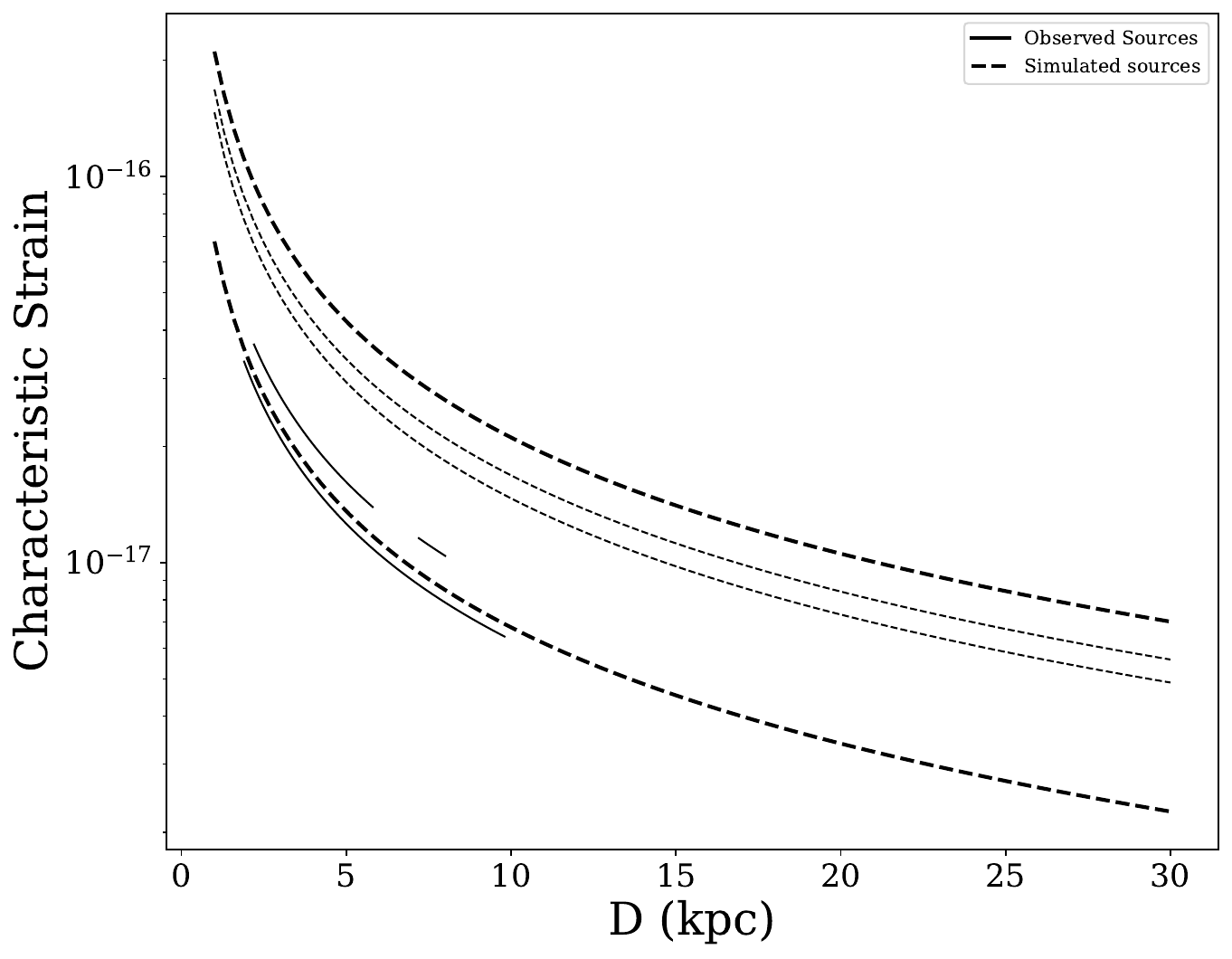}
\caption{Characteristic strain as a function of distance for observed and simulated UCXB sources. The solid lines represent observed sources, while the dashed lines show simulated sources.}
\label{fig:5}
\end{figure}

Using the LEGWORK package, we calculated the characteristic strain for 48 simulated UCXB systems. In Figure \ref{fig:5}, we plot the upper and lower bounds of the characteristic strain as a function of distance, covering the range from 1 to 30 kpc. This plot illustrates how the characteristic strain systematically decreases as the distance from the source increases, providing a clear depiction of the expected GW emissions from UCXBs. The dashed lines in the figure represent the upper and lower bounds of the characteristic strain for the simulated UCXBs, which gives a comprehensive view of the emission profile for different systems.

Additionally, we utilized the data from Table \ref{tab:1} to calculate the chirp mass and characteristic strain for the selected UCXBs and then overlaid these calculated characteristic strains onto the simulated data. The solid lines represent these observed sources, and the range of characteristic strains accounts for uncertainties in their measured distances \citep{2016A&A...595A...1G}. By plotting these observed systems alongside the simulated ones, we can directly compare their characteristic strains. This comparison highlights the alignment between real observations and our simulation results.

The fact that the observed sources fall within the range of characteristic strains predicted by our simulations confirm that the model \textbf{may} captures the key physical processes underlying UCXB evolution and GW emission, making it a reliable tool for predicting the GW characteristics and validating its application for future studies. This reinforces the notion that UCXBs are stable and detectable sources of GWs, both in terms of their long-term evolution and their GW emissions across a range of distances.

\subsubsection{The Sensitivity Curve and the Population of Sources}

In this study, we conducted GW signal simulations for simulated UCXBs as well as observed UCXBs, which are likely formed through the MS channel, using the LEGWORK tool to analyze these systems. Our simulated UCXB sources distributed randomly within a distance range of 2-12 kpc, reflecting realistic observational uncertainties and typical distances of known UCXBs.

We then evaluated their detectability with the space-based GW detectors Taiji and TianQin. Our research focuses on the performance of the detectors over different observation periods (short-term: 1 year, medium-term: 4 years, and long-term: 8 years) and analyzes their potential to detect UCXBs, particularly in terms of frequency and SNR.

The ASD is a measure of the noise in a GW detector as a function of frequency. It is derived from the noise power spectral density (PSD), $S_n(f)$, where $f$ is the frequency. The ASD is defined as the square root of the PSD:

\begin{equation}
    \text{ASD}(f) = \sqrt{S_n(f)} \, [\text{Hz}^{-1/2}]
\end{equation}

The ASD provides a convenient way to visualize the sensitivity of a detector across different frequency bands. A lower ASD value corresponds to a more sensitive detector at a given frequency. In gravitational wave astronomy, the ASD is commonly used to compare the sensitivity of detectors.

We compared the performance of Taiji and TianQin in detecting UCXBs across different frequency bands and identified their respective advantages and limitations. Taiji exhibits superior detectability in the low-frequency range (approximately \(10^{-4} \, \text{Hz} \) to \(10^{-2} \, \text{Hz}\)), while TianQin's strength lies in the high-frequency range (approximately \(10^{-2} \, \text{Hz}\) to \(1 \, \text{Hz}\)). However, considering the frequency distribution of UCXBs, they are primarily concentrated in Taiji's optimal detection range (\(10^{-4} \, \text{Hz}\) to \(10^{-2} \, \text{Hz}\)), which makes Taiji more favorable for UCXB detection overall.

By comparing the simulated UCXBs systems with observed UCXBs, we validated the reliability of the simulations.Our results indicate that only systems with $\mathrm{SNR} > 5$ are likely to be detected reliably. Although prior studies such as those of \citet{2020ApJ...900L...8C} provided valuable insight into detectability by focusing on the characteristic strain exceeding the detector sensitivity curve, \textbf{this approach establishes an upper limit  for the detection reliability of GW signals.} By incorporating SNR into our analysis using LEGWORK, we improve the evaluation of detectable sources and emphasize the importance of extended observation durations in improving both sensitivity and detection confidence.

The simulated GW signals are in good agreement with the observed UCXBs, demonstrating that our models \textbf{may} predict UCXBs that future space-based detectors could observe.

\begin{figure*}
\centering
\includegraphics[width=2\columnwidth]{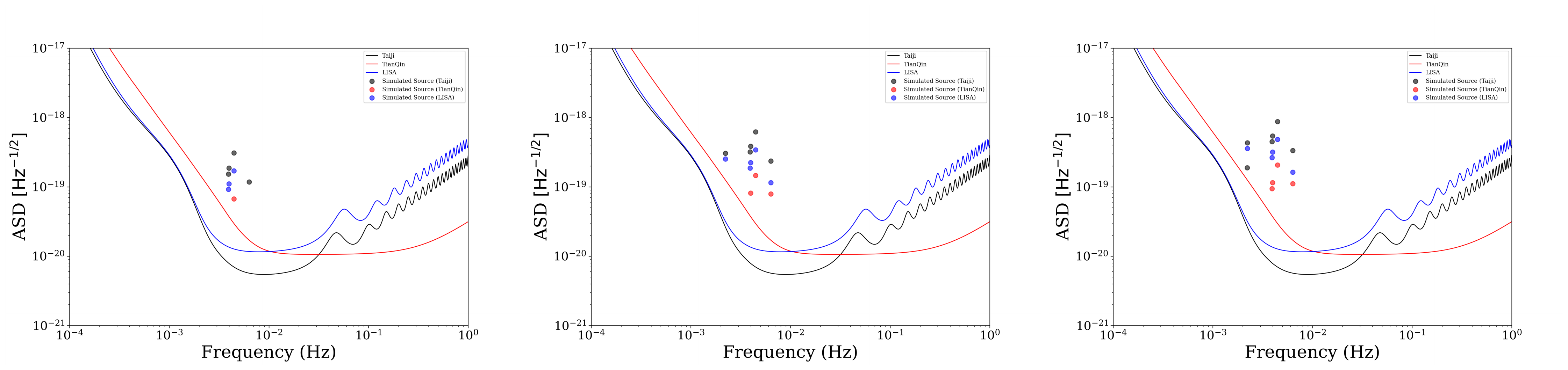}
\caption{ASD as a function of frequency for both simulated and observed UCXB systems, overlaid with the sensitivity curves of Taji (black line) and TianQin (red line). From left to right, the panels depict results for observation periods of 1 year, 4 years, and 8 years. The circles represent simulated UCXB sources.}
\label{fig:6}
\end{figure*}

As illustrated in Figure \ref{fig:6}, LISA and Taiji show remarkable sensitivity in the low-frequency range, making them capable of detecting more low-frequency UCXBs. Over longer observation periods, \textbf{their ability} to detect weaker and more distant sources improves significantly. This detection confirms LISA and Taiji's observational capabilities, particularly in identifying UCXBs in the low-frequency range, and underscores the reliability of its performance over extended observation periods. TianQin is also capable of detecting simulated UCXBs and can successfully capture these sources over extended observation periods. These results highlight the critical role of extended observation times in improving the sensitivity of space-based detectors and their ability to capture a greater number of simulated UCXB sources, thereby deepening our understanding of their GW signals. Thus, TianQin’s contributions to joint observations are vital for constructing a more comprehensive picture of the UCXB population and improving the overall understanding of their GW emissions.

\begin{figure}
\centering
\includegraphics[width=1\columnwidth]{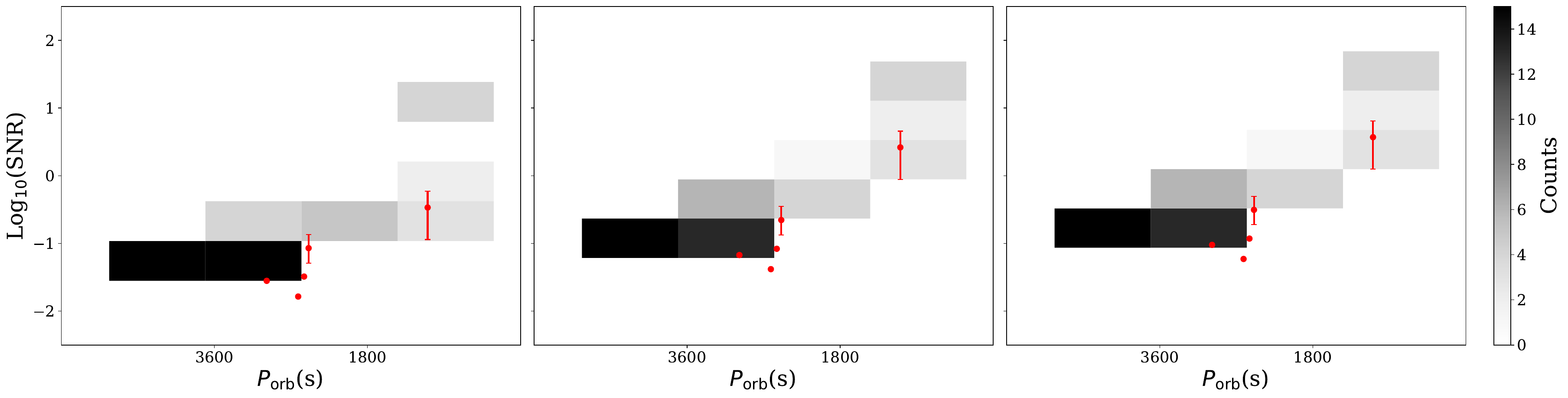}
\caption{Distribution of SNR in logarithmic scale (\( \log_{10}(\text{SNR}) \)) as a function of orbital period by Taiji. The grayscale shading indicates the count of sources in each bin, with darker shades representing higher counts. Red points represent real UCXBs, and error bars indicate the uncertainty in their measurements. Panels from left to right correspond to observation periods of 1 year, 4 years, and 8 years.}
\label{fig:7}
\end{figure}

Figure \ref{fig:7} illustrates the detectability of UCXB systems by Taiji, showing the distribution of sources across orbital periods and their corresponding SNRs. The heatmap highlights that most simulated sources fall within Taiji’s sensitivity range, particularly for systems with shorter orbital periods (below 30 minutes) and higher SNRs. The gradual increase in source counts as SNR rises reflects Taiji’s strong ability to detect a substantial portion of the UCXB population, particularly in the low-frequency regime where these systems predominantly reside.

The red dots represent observed UCXBs, and the error bars associated with these points originate from uncertainties in distance measurements \citep{2016A&A...595A...1G}. These uncertainties directly affect the derived SNR, as the SNR depends on the inverse square of the distance to the source. Despite these uncertainties, the alignment of observed UCXBs with the simulated source distribution demonstrates the reliability of the models in capturing the characteristics of  UCXBs. Taiji’s ability to encompass these systems within its sensitivity range underscores its potential as a highly capable instrument for studying UCXBs and their GW emissions.

\begin{figure}
\centering
\includegraphics[width=\columnwidth]{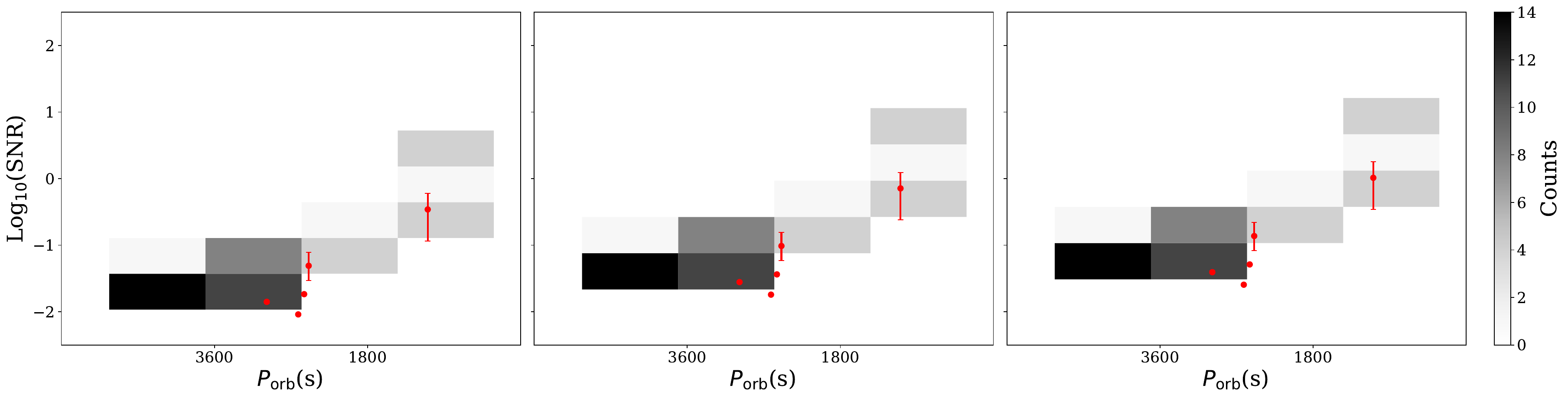}
\caption{Similar to Fig. 7, we present the distribution of SNR in logarithmic scale (\( \log_{10}(\text{SNR}) \)) as a function of orbital period for the TianQin.}
\label{fig:8}
\end{figure}

The Figure \ref{fig:8} illustrates the detectability of UCXB systems by TianQin. The heatmap shows that TianQin generally achieves lower SNRs compared to Taiji, but achieves its highest sensitivity for UCXBs with extremely short orbital periods. These systems are located in the high-frequency range where TianQin excels, making it especially suitable for detecting such compact and fast-evolving sources.

By combining the observations from LISA, Taiji and TianQin, joint data provides a significantly more reliable and robust approach to detecting UCXBs. Each detector has its own unique noise characteristics and sensitivity profiles, and cross-referencing data from different instruments greatly improves the precision of the detections. This synergy not only enhances the overall reliability of our findings but also ensures that UCXBs are identified with higher accuracy and confidence.

More importantly, joint observations enable the precise localization of UCXBs by exploiting the timing differences in the signals received by the detectors. Given that LISA, Taiji and TianQin are positioned at different locations in space and are sensitive to distinct portions of the frequency spectrum, combining their data allows for the triangulation of the source's position with much greater certainty. This cross-checking across detectors and frequency ranges helps to refine the sky map, reducing the uncertainty in source localization.

By leveraging the complementary strengths of the detectors, joint observations not only enhance detection accuracy but also play a pivotal role in improving the precision of UCXB localization. This refined localization ensures that sources are pinpointed with greater confidence, even when their signals are faint or distant.

We roughly calculated the combined SNR for joint observations, using the following formula:
\begin{equation}
\text{SNR}_{\text{combined}} = \sqrt{\text{SNR}_{\text{1}}^2 + \text{SNR}_{\text{2}}^2 }
\end{equation}

This equation assumes that the noise contributions of each detector are uncorrelated, providing a straightforward way to calculate the enhancement in SNR through joint observations. The results demonstrate significant improvements in detection confidence when combining data from multiple detectors, as summarized in Table \ref{tab:2}.

Besides the increase in the observable number of simulated UCXBs, we emphasize that the well-known source 4U 1543-624 achieves a combined SNR greater than 5 under long-term observations by Taiji and LISA. This result strongly supports the importance of joint observations, highlighting their significance for studying actual UCXBs.

\begin{table*}
\centering
\caption{Counts of simulated UCXBs detected (\(\text{SNR} > 5\)) for detectors and their combinations}
\begin{tabular}{l c c c c c c c}
\hline
\textbf{Observation Time} & \textbf{TianQin} & \textbf{LISA} & \textbf{Taiji} & \textbf{LISA + TianQin} & \textbf{Taiji + TianQin} & \textbf{Taiji + LISA} & \textbf{All Detectors} \\
\hline
1 Year & 1 & 3 & 4 & 4 & 4 & 4 & 4 \\
4 Years & 3 & 5 & 5 & 5 & 5 & 6 & 6 \\
8 Years & 4 & 5 & 6 & 6 & 6 & 7 & 7 \\
\hline
\end{tabular}
\label{tab:2}
\end{table*}

\subsection{Estimating the Current Observable Population of UCXBs From Low Mass Main-Sequence Companion Channel}

To estimate the current observable number of UCXBs from the MS companion channel in the Milky Way, we focused on binary systems satisfying specific initial conditions: \(M_1 = 8{-}12 \, M_\odot\), \(M_2 = 0.5{-}1.2 \, M_\odot\), \(R = 3{-}9 \, \mathrm{AU}\), and \(Z = 0.002{-}0.01\), as shown in Figure \ref{fig:2}. Note that in our simulations, we assumed that these stars form simultaneously, which is a simplifying assumption for this study. From our simulations, this parameter space contained 339,669 binaries, within which 48 UCXBs formed, yielding a formation rate:
\begin{equation}
  P_{\text{UCXB|Simulated}} = \frac{48}{339,669} \approx 1.41 \times 10^{-4}.
\end{equation}

The total number of binaries in the Milky Way is estimated as \citep{2015ApJ...806...96L}:
\begin{equation}
  N_{\text{binary}} = (5 \times 10^{10} \text{ to } 2.4 \times 10^{11}),
\end{equation}
The initial mass function (IMF) follows the Kroupa form \citep{2001MNRAS.322..231K,2003PASP..115..763C}:
\begin{equation}
  \xi(M) \propto M^{-\alpha}, \quad \alpha =
  \begin{cases} 
    1.3, & \text{if } M < 0.5 \, M_\odot,  \\ 
    2.3, & \text{if } M > 0.5 \, M_\odot.
  \end{cases}
\end{equation}

The combined fraction \(f_{\text{Initial}}\) is the product of mass fraction \(f_{\text{mass}}\), orbital radius fraction \(f_R\) \citep{2013ARA&A..51..269D}, and metallicity fraction \(f_Z\). These fractions are calculated as follows:
\begin{equation}
  f_{\text{mass}} \approx 0.004, \quad f_R \approx 0.12, \quad f_Z \approx 0.20,
\end{equation}
so the total initial fraction is:
\begin{equation}
  f_{\text{Initial}} = f_{\text{mass}} \times f_R \times f_Z = 0.004 \times 0.12 \times 0.20 = 0.000096.
\end{equation}

The total number of UCXBs from the MS companion channel formed over Galactic history is:
\begin{equation}
  \begin{split}
    N_{\text{total}} &= f_{\text{Initial}} \times N_{\text{binary}} \times P_{\text{UCXB|Simulated}} \\
    &= (676 \text{ to } 3243).
  \end{split}
\end{equation}

A typical UCXB from this channel has a lifetime of \(100 \, \mathrm{Myr}\) \citep{2021MNRAS.506.4654W} and a Galactic star formation history of \(10 \, \mathrm{Gyr}\). We assume that UCXBs are formed uniformly over time, meaning the formation rate of UCXBs is constant at all time points throughout the evolution of the Milky Way. Based on this, the fraction currently observable is:
\begin{equation}
  \frac{\text{lifetime}}{\text{Galactic star formation duration}} = 0.01.
\end{equation}
Thus, the current observable number is:
\begin{equation}
  N_{\text{current}} = N_{\text{total}} \times 0.01 = (7 \text{ to } 32).
\end{equation}

In summary, our simulations suggest that the Milky Way currently hosts approximately \(7{-}32\) observable UCXBs from the MS companion channel, depending on the total binary population and initial parameter distributions. This range is broadly consistent with the 5 selected UCXBs that have been observed to date. The agreement between our theoretical predictions and the observed sample strengthens the reliability of the simulations and validates the parameter space selected for the formation of UCXB from the MS companion channel.

\section{Conclusion} 
\label{sec:cite}

This study provides a comprehensive analysis of UCXB formation and evolution using the COMPAS binary evolution code, starting from the ZAMS. While the COMPAS code relies on certain approximations and simplifications, which means the study may not capture all aspects of UCXB evolution in a fully comprehensive manner, it still generates results that are consistent with the known properties of UCXBs. By modeling the entire evolutionary history of UCXBs, rather than focusing solely on specific stages RLOF or CE phases, we explore a broader range of evolution. Although the study is not exhaustive, it provides a solid foundation for understanding UCXB evolution and produces results that align well with the observed characteristics of UCXBs.

Our study also advances the detectability analysis of UCXBs as GW sources by incorporating SNR calculations into population synthesis models. Although previous studies, such as those by Chen et al. \citep{2020ApJ...900L...8C}, relied mainly on characteristic strain comparisons to detector sensitivity curves, our approach evaluates the detection reliability more realistically. We specifically compute SNR values for different observation durations and apply a threshold of $\text{SNR} > 5$ to provide a robust framework for assessing the detectability of UCXBs, particularly for observations with LISA, Taiji and TianQin. The alignment of our simulated GW signals with the ASD curves of these detectors validates the feasibility of detecting UCXBs formed through MS donor channels, even at distances up to 12 kpc. \textbf{These results may indicate the robustness of our simulations and could potentially highlight the importance of extended observation durations in improving the detection of distant and marginal sources.}

In addition, our analysis estimates that the Milky Way currently hosts approximately \(7{-}32\) observable UCXBs from the MS companion channel. This range is broadly consistent with the observed data so far, suggesting that our simulations and selected parameter space \textbf{may} reflect the real population of UCXBs from the MS companion channel in the Galaxy. The agreement between theoretical predictions and observational data not only strengthens the validity of our modeling framework, but also highlights the potential of space-based GW observatories to identify more UCXBs in the future.

Based on our simulations, Taiji or LISA alone could detect \(1{-}6\) UCXBs from the MS companion channel in the Milky Way over a long-term observation period, while TianQin could detect \(1{-}4\) sources. Joint observations with TianQin and LISA could enhance this number, improve localization, and contribute to a deeper understanding of UCXBs as GW sources. This underscores the importance of both individual and collaborative observations for uncovering the galactic population of these UCXBs.

Furthermore, due to the significant influence of mass on GW signals, we argue that space-based GW detectors such as Taiji and LISA are well-suited to capturing BH-UCXBs. The inclusion of BH-UCXBs as potential sources expands the scope of GW astronomy, highlighting the capability of these detectors to explore both classical and less-studied compact binary populations.

Our findings confirm that UCXBs, particularly those with extreme orbital periods, are significant sources of GWs within the sensitivity ranges of next-generation space-based detectors. The combination of stable GW emission over long timescales and intense signals during late-stage orbital contraction underscores their importance as key targets for future observations. By employing population synthesis, we provide a detailed and realistic view of UCXB formation and evolution, emphasizing the importance of modeling the entire evolutionary path rather than focusing on individual phases. This approach enables the identification of new UCXB formation pathways and contributes to a deeper understanding of their role in the broader context of compact binary evolution.

In conclusion, this work significantly strengthens the case for UCXBs as important sources in multi-messenger astronomy. By providing a comprehensive framework for their formation, evolution, and detectability, it lays the groundwork for future observational missions and theoretical studies. 

\section*{\centering Acknowledgments}

This work was supported by the Tianshan Talent Training  Program through the grant 2023TSYCCX0101, the Central Guidance for Local Science and Technology Development Fund under No. ZYYD2025QY27, the  National Natural Science Foundation of China NSFC 12433007 and the science research grants from the China Manned Space Project with No. CMS-CSST-2021-A08.


\bibliography{UCXB_Space_based_GW_detector_}{}
\bibliographystyle{aasjournal}



\end{document}